\documentclass[letter]{aa} 
\usepackage{graphics}
\usepackage{graphicx}
\usepackage{multicol}
\usepackage[varg]{txfonts}

\usepackage{natbib,twoopt}
\usepackage{amsmath} 
\usepackage[breaklinks=true]{hyperref} 
\bibpunct{(}{)}{;}{a}{}{,} 
\makeatletter
\newcommandtwoopt{\citeads}[3][][]{\href{http://adsabs.harvard.edu/abs/#3}%
{\def\hyper@linkstart##1##2{}%
\let\hyper@linkend\@empty\citealp[#1][#2]{#3}}}
\newcommandtwoopt{\citepads}[3][][]{\href{http://adsabs.harvard.edu/abs/#3}%
{\def\hyper@linkstart##1##2{}%
\let\hyper@linkend\@empty\citep[#1][#2]{#3}}}
\newcommandtwoopt{\citetads}[3][][]{\href{http://adsabs.harvard.edu/abs/#3}%
{\def\hyper@linkstart##1##2{}%
\let\hyper@linkend\@empty\citet[#1][#2]{#3}}}
\newcommandtwoopt{\citeyearads}[3][][]%
{\href{http://adsabs.harvard.edu/abs/#3}
{\def\hyper@linkstart##1##2{}%
\let\hyper@linkend\@empty\citeyear[#1][#2]{#3}}}
\makeatother

\usepackage{color}
\definecolor{mygreen}{RGB}{0,128,0}

\hypersetup{colorlinks=true,linkcolor=blue,citecolor=blue,urlcolor=blue}

\def\barypmRA{-3619.9} 
\def\barypmdec{+693.8} 
\def\barypmerr{3.9} 
\def\ProxpmRA{-3773.8} 
\def\ProxpmRAerr{0.4} 
\def\Proxpmdec{+770.5} 
\def\Proxpmdecerr{2.0} 
\def\radiusA{1.2234}

\def\radiusAerrtot{0.0053} 

\def\radiusB{0.8632}

\def\radiusBerrtot{0.0037} 

\def\massA{1.1055}
\def\massAerr{0.0039}

\def\massB{0.9373}
\def\massBerr{0.0033}

\def\massAB{2.0429}
\def\massABerr{0.0072}
\def\mProx{0.123}
\def\mProxerr{0.006}
\def\mProxMann{0.1221}
\def\mProxMannerr{0.0022}
\def\RProx{0.141}
\def\RProxerr{0.007}
\def\RProxMann{0.1542}
\def\RProxMannerr{0.0045}

\def\parallaxAlfCen{747.17}
\def\parallaxAlfCenerr{0.61}
\def\parallaxProx{768.77}
\def\parallaxProxerr{0.37}
\def\distABProx{12\,947}
\def\distABProxerr{260}
\def\GRSAlfCen{+575.3}
\def\GRSAlfCenerr{2.4}
\def\GRSProx{+504}
\def\GRSProxerr{17}
\def\dGRSAlfCen{-61.4}
\def\dGRSAlfCenerr{2.7}
\def\RVAlfCenObs{-22.393}
\def\RVAlfCenObserr{0.004}
\def\RVProxObs{-21.700}
\def\RVProxObserrstat{0.011} 
\def\RVProxObserr{0.027}
\def\RVAlfCen{-22.332}
\def\RVAlfCenerr{0.005}
\def\RVProx{-22.204}
\def\RVProxerr{0.032}
\def\velprox{273 \pm 49}
\def\sma{8.7^{+0.7}_{-0.4}} 
\def\ecc{0.50^{+0.08}_{-0.09}}
\def\Per{547^{+66}_{-40}}
\def\incl{107.6^{+1.8}_{-2.0}}
\def\Omegaval{126^{+5}_{-5}}
\def\omegaval{72.3^{+8.7}_{-6.6}}
\def\Tzero{+283^{+59}_{-41}}
\def\Periastron{4.3^{+1.1}_{-0.9}}
\def\Apastron{13.0^{+0.3}_{-0.1}}

\begin{document} 

\title{Proxima's orbit around $\alpha$\,Centauri\thanks{Based on observations collected at the European Organisation for Astronomical Research in the Southern Hemisphere under ESO programs 072.C-0488(E), 082.C-0718(B), 183.C-0437(A), 191.C-0505(A) and 096.C-0082(A).}}
\titlerunning{Proxima's orbit around $\alpha$\,Centauri}
\authorrunning{P. Kervella et al.}

\author{
P.~Kervella\inst{1,2}
\and
F.~Th\'evenin\inst{3}
\and
C.~Lovis\inst{4}
}
\institute{
Unidad Mixta Internacional Franco-Chilena de Astronom\'{i}a (CNRS UMI 3386), Departamento de Astronom\'{i}a, Universidad de Chile, Camino El Observatorio 1515, Las Condes, Santiago, Chile, \email{pkervell@das.uchile.cl}.
\and
LESIA (UMR 8109), Observatoire de Paris, PSL Research University, CNRS, UPMC, Univ. Paris-Diderot, 5 Place Jules Janssen, 92195 Meudon, France, \email{pierre.kervella@obspm.fr}.
\and
Universit\'e C\^ote d'Azur, Observatoire de la C\^ote d'Azur, CNRS, Lagrange UMR 7293, CS 34229, 06304, Nice Cedex 4, France.
\and
Observatoire astronomique de l'Universit\'e de Gen\`eve, 51 Ch. des Maillettes, 1290 Versoix, Switzerland
}
\date{Received ; Accepted}

 
  \abstract
{
Proxima and $\alpha$\,Centauri AB have almost identical distances and proper motions with respect to the Sun.
Although the probability of such similar parameters is in principle very low, the question as to whether they actually form a single gravitationally bound triple system has been open since the discovery of Proxima one century ago.
Owing to HARPS high precision absolute radial velocity measurements and the recent revision of the parameters of the $\alpha$\,Cen pair, we show that Proxima and $\alpha$\,Cen are gravitationally bound with a high degree of confidence.
The orbital period of Proxima is approximately 550\,000 years.
With an excentricity of $\ecc$, Proxima comes within $\Periastron$\,kau of $\alpha$\,Cen at periastron.
Its orbital phase is currently close to apastron ($\Apastron$\,kau).
This orbital motion may have influenced the formation or evolution of the recently discovered planet orbiting Proxima as well as circumbinary planet formation around $\alpha$\,Cen.
}

   \keywords{Stars: individual: $\alpha$ Cen; Stars: individual: Proxima; Stars: binaries: visual; Astrometry; Proper motions; Celestial mechanics}

   \maketitle
%

 
\section{Introduction} 
 
The visual triple star comprising $\alpha$ Centauri (\object{WDS J14396-6050AB}, \object{GJ559AB}) and Proxima (\object{HIP 70890}, \object{GJ551}) is the nearest stellar system to the Earth.
The solar-like $\alpha$\,Cen A (spectral type G2V, \object{HD 128620}) and the cooler dwarf $\alpha$\,Cen B (\object{HD 128621}) are located at a distance of only $d = 1.3384 \pm 0.0011$\,pc \citepads{2016kervella}.
The third star \object{Proxima} is a cool red dwarf (M5.5V), that is closer to Earth by approximately 7\,800\,au, at $d = 1.3008 \pm 0.0006$\,pc \citepads{1999AJ....118.1086B}.
Owing to their similarity to the Sun, $\alpha$\,Cen A and B are benchmarks for both stellar physics \citepads{2016MNRAS.460.1254B} and extrasolar planet research \citepads{2015MNRAS.450.2043D}.
In August 2016, \citetads{2016Natur.536..437A} announced the discovery of a terrestrial-mass planet orbiting Proxima in its habitable zone (\object{Proxima b}).
The presence of a potentially life-sustaining planet around our nearest stellar neighbor is a strong incentive for the \emph{Breakthrough Starshot}\footnote{\url{https://breakthroughinitiatives.org/Initiative/3}} initiative to send ultra-fast light-driven nanocrafts to $\alpha$\,Centauri. 

Proxima was discovered more than one century ago by \citetads{1915CiUO...30..235I}, and the strong similarity of its proper motion and parallax with those of $\alpha$\,Cen was immediately noticed \citepads{1926CiUO...70..390I,1925BHarO.818....2L,1928AJ.....39...20A}.
The question as to whether Proxima is gravitationally bound to $\alpha$\,Cen has been discussed by several authors (\citeads{1966AJ.....71.1017G}; \citeads{1979ApJ...234L.205W}; \citeads{1993MNRAS.261L...5M}; \citeads{1994A&A...292..115A}; \citeads{2006AJ....132.1995W}).
Although statistical considerations are usually invoked to justify that Proxima is probably in a bound state, solid proof from dynamical arguments using astrometric and radial velocity (RV) measurements have never been obtained at a sufficient statistical significance level.
As discussed by \citetads{2016arXiv160703090W}, if Proxima is indeed bound, its presence may have impacted planet formation around the main binary system.

\section{Radial velocities\label{radvel}}

\subsection{Observed radial velocities}

We adopt the observed RV of the barycenter of $\alpha$\,Cen A and B determined by \citetads{2016kervella} that is statistically identical to the value obtained by \citetads{2016A&A...586A..90P} from the same RV dataset
$v_\mathrm{r,\,obs}[\mathrm{\alpha\,Cen}] = \RVAlfCenObs \pm \RVAlfCenObserr\,\mathrm{km\,s}^{-1}$.

The main obstacle in demonstrating that Proxima is gravitationally bound to $\alpha$\,Cen has historically been the lack of very-high-precision RVs of Proxima (see e.g.,~\citeads{1967Obs....87...79T}).
This is a consequence of its relative faintness in the visible ($m_V = 11$), but the exquisite accuracy and sensitivity achieved by modern planet-search spectrographs has overcome this limitation.
We considered the possibility of adopting the RV of Proxima $v_\mathrm{r,\,obs}[\mathrm{Proxima}]  = -22.345\,\mathrm{km\,s}^{-1}$ published by \citetads{2014MNRAS.439.3094B}.
However, the method they used to absolutely reference the velocity of their mask (\object{GJ 1061}) is uncertain.
While the differential velocity between the mask and Proxima is  measured with an accuracy of a few meters per second or better, the absolute value may be affected by large errors that could reach several hundred meters per second.
In order to obtain the absolute velocity of Proxima, we thus went back to the original HARPS spectra \citepads{2006SPIE.6269E..0PL} obtained  between 2004 and 2016.
The details on our measurement technique are provided in Appendix~\ref{radvelProxima}.
We obtain a RV of $v_\mathrm{r,\,obs}[\mathrm{Proxima}] = \RVProxObs \pm \RVProxObserr\,\mathrm{km\,s}^{-1}$.
The RV variations induced by \object{Proxima b} are negligible (1.38\,m\,s$^{-1}$; \citeads{2016Natur.536..437A}).

\subsection{Convective blueshift}

The convective blueshift (CB) is a systematic displacement of the wavelengths of a star's spectral lines that is induced by the structure of its surface convection pattern (\citeads{1982ARA&A..20...61D}; \citeads{2011ApJ...733...30S}).
The upward moving material in convective cells usually occupies a larger surface and is hotter than the downward moving gas in the intergranular lanes.
The net result is a systematic displacement of the spectral lines forming close to the photosphere of the star, in general toward the blue (but not systematically), that is, a negative RV shift.
The CB is stronger for hotter stars with convective surfaces, and for larger convective cells \citepads{2008sf2a.conf....3B}.

The template that was used by \citetads{2016A&A...586A..90P} for the cross-correlation of the spectra of $\alpha$\,Cen~A and the derivation of the RV measurement is the Fourier transform spectrum of the Sun from \citetads{1984sfat.book.....K}. 
The resulting RVs were corrected by \citetads{2016A&A...586A..90P} for the zero point determined by \citetads{2013Msngr.153...22M}.
The accuracy of this calibration was confirmed at a 2\,m\,s$^{-1}$ level by \citetads{2016MNRAS.457.3637H}, so it is extremely solid.
The solar spectrum is an excellent match to the spectrum of $\alpha$\,Cen~A and the cross correlation therefore automatically takes the CB into account.
Their effective temperature $T_\mathrm{eff}$, that is, the flux emitted per unit surface of the photosphere, is identical to less than 20\,K \citepads{2016bkervella}.
Their effective gravity $\log g$ is also very close: $\log g [\alpha\,\mathrm{Cen\,A}] = 4.3117 \pm 0.0015$ \citepads{2016bkervella} whereas $\log g [\odot] = 4.4384$, as the larger radius of $\alpha$\,Cen\,A compensates for its higher mass. 
The surface convection thus operates in essentially identical conditions, and its properties in both stars are expected to be very similar.
This similarity is essential to be insensitive to the CB uncertainty and to reach the highest absolute RV accuracy \citepads{2008A&A...492..841R}.
As a remark, \citetads{2002A&A...386..280P} measured a CB difference of only $72 \pm 26$\,m\,s$^{-1}$ between $\alpha$\,Cen~A and B, although they have significantly different $T_\mathrm{eff}$ and $\log g$.
The difference in CB between the Sun and $\alpha$\,Cen~A is thus probably one order of magnitude smaller.
We therefore neglect the difference in CB between the Sun ($\approx -300$\,m\,s$^{-1}$; \citeads{1999ASPC..185..268D}) and $\alpha$\,Cen~A.

The CB and gravitational redshift (GRS, see Sect.~\ref{gravitationalredshift}) of $\alpha$\,Cen~B are significantly different from the Sun.
Their combined effect is taken into account in the orbital fits by \citetads{2016kervella} and \citetads{2016A&A...586A..90P}, through a constant differential velocity term $\Delta V_B$ between $\alpha$\,Cen A and B.
Its amplitude is estimated to $\Delta V_B = 322 \pm 5$\,m\,s$^{-1}$ \citepads{2016kervella}.
This correction term $\Delta V_B$ also compensates for the mismatch of the template spectral mask.
A small constant term is also adjusted for A ($\Delta V_A = 8 \pm 5$\,m\,s$^{-1}$) by \citetads{2016kervella} to obtain a better quality fit of the full data set including astrometry, but it is marginally significant.
In summary, the corrective term $\Delta V_B$ brings the RV of $\alpha$\,Cen~B into the same barycentric referential as $\alpha$\,Cen~A, that is itself securely an absolute velocity thanks to the solar template used for the cross-correlation.
This translates into an absolutely calibrated RV for the barycenter of $\alpha$\,Cen~A and B, that is by construction insensitive to the CB of both stars.
However, we still have to correct the RV of the barycenter for the differential gravitational redshift ($\Delta v_\mathrm{GRS}$) of $\alpha$\,Cen A with respect to the Sun (Sect.~\ref{gravitationalredshift}).

For Proxima, the small expected size of the convective cells results in a very small predicted CB.
In addition, our choice to measure its RV using emission lines that form essentially in the chromosphere is naturally less sensitive to CB.
In the red dwarf Barnard's star (\object{GJ 699}), whose properties are similar to Proxima, \citetads{2003A&A...403.1077K} proposed that the effect of convection is actually a redshift and not a blueshift.
Through an analogy with the Sun, they estimate an upper limit of +33\,m\,s$^{-1}$, and variability with the magnetic field strength and location of stellar spots.
The presence of magnetic field is likely to inhibit the convective flow \citepads{2016A&A...593A.127K}, hence affecting the convective pattern at the surface.
But the velocity of Proxima is stable at a level of a few m\,s$^{-1}$ over long periods (\citeads{2014MNRAS.439.3094B}; \citeads{2016Natur.536..437A}).
Owing to this stability, we neglect the effect of CB in Proxima's RV with respect to the other uncertainties.

\subsection{Gravitational redshift\label{gravitationalredshift}}

According to general relativity, the wavelength of the photons emitted by a star are shifted to the red as they climb out of its gravitational well.
This results in a shift of the wavelength of the spectral lines toward the red (i.e., longer wavelengths).
The GRS effect on the RV is a function of the mass $m$ of the star and its radius $R$  through $v_\mathrm{GRS} = G\,m / (R\,c)$.
The GRS has been observed by \citetads{2012SoPh..281..551T} in the Sun with an amplitude comparable with the expected value of $+633$\,m\,s$^{-1}$, but its detection in main sequence and giant stars of the M67 stellar cluster remains elusive \citepads{2011A&A...526A.127P}.
As it is a function of $m/R$, compact objects (white dwarfs, neutron stars, and black holes) create the strongest GRS, typically +40 km\,s$^{-1}$ for white dwarfs \citepads{2012ApJ...757..116F}, while giants and supergiants exhibit very small GRS \citepads{1999ASPC..185..268D}.

As the RV of $\alpha$\,Cen A was determined by \citetads{2016A&A...586A..90P} from cross-correlation with a solar spectrum template, the GRS of the Sun is incorporated in the derived RV values (the velocity of the Sun is zero when the solar template is cross-correlated with a solar spectrum).
However, we have to include a differential GRS term:
\begin{equation}
\Delta v_\mathrm{GRS}(\alpha\,\mathrm{Cen\,A}) = \mathrm{GRS}(\alpha\,\mathrm{Cen\,A}) - \mathrm{GRS}(\mathrm{Sun}).
\end{equation}
To estimate the GRS of $\alpha$\,Cen\,A, its mass is taken from \citetads{2016kervella} and its radius from \citetads{2016bkervella} (Table~\ref{physicalparameters}).
We obtain $v_\mathrm{GRS}[\mathrm{\alpha\,Cen\,A}] = \GRSAlfCen \pm \GRSAlfCenerr$\,m\,s$^{-1}$.
Subtracting the GRS of the Sun gives $\Delta v_\mathrm{GRS}(\alpha\,\mathrm{Cen\,A}) = \dGRSAlfCen \pm \dGRSAlfCenerr$\,m\,s$^{-1}$.
The $\Delta v_\mathrm{GRS}$ corrected, absolute velocity of the barycenter of $\alpha$\,Cen is therefore $v_\mathrm{r,\,abs}[\mathrm{\alpha\,Cen}] = \RVAlfCen \pm \RVAlfCenerr\,\mathrm{km\,s}^{-1}$.

We estimated the RV of Proxima from a direct comparison of the wavelengths of its emission lines to their laboratory wavelengths.
So unlike for $\alpha$\,Cen~A, we have here to correct for the full amplitude of the GRS ($v_\mathrm{GRS}$).
The mass of Proxima is not directly measured.
\citetads{0004-637X-804-1-64} (see also \citeads{2016ApJ...819...87M}) used a large sample of M and K dwarfs to calibrate polynomial relations between the absolute $K_s$ magnitude and the mass or the radius.
The 2MASS $K_s$ magnitude of Proxima is $m_{Ks} = 4.384 \pm 0.033$ \citepads{2003yCat.2246....0C}, corresponding to an absolute magnitude $M_{Ks} = 8.813 \pm 0.033$.
Using the mass-$M_{Ks}$ relation from \citetads{0004-637X-804-1-64}, we obtain $m_\mathrm{Prox} = \mProxMann \pm \mProxMannerr\,M_\odot$.
The derived mass is in perfect agreement with the value of $m_\mathrm{Prox} = \mProx \pm \mProxerr\,M_\odot$ resulting from the mass-luminosity relation by \citetads{2000A&A...364..217D}.
The radius-$M_{Ks}$ relation gives $R_\mathrm{Prox} = \RProxMann \pm \RProxMannerr\,R_\odot$, which is slightly larger ($+1.6\sigma$) than the interferometrically measured value of $R_\mathrm{Prox} = \RProx \pm \RProxerr\,R_\odot$ from \citetads{2009A&A...505..205D}.
We adopt the radius predicted by the relation from \citetads{0004-637X-804-1-64} as it is determined using the same underlying star sample as was the mass.
In addition to being more precise than the interferometric measurement, this choice reduces the potential systematics on $m/R$, that is the quantity of interest to determine the GRS.
We obtain a GRS of $v_\mathrm{GRS}[\mathrm{Proxima}] = \GRSProx \pm \GRSProxerr$\,m\,s$^{-1}$.
The GRS is an important source of uncertainty on the RV of Proxima, and it is also a significant contributor for $\alpha$\,Cen AB.
We apply this correction to the measured RVs of Proxima and obtain 
$v_\mathrm{r,\,abs}[\mathrm{Proxima}]  = \RVProx \pm \RVProxerr\,\mathrm{km\,s}^{-1}$.

\begin{table}[ht]
        \caption{Adopted physical parameters of $\alpha$\,Cen AB and Proxima.}
        \label{physicalparameters}
        \centering
        \renewcommand{\arraystretch}{1.2}
        \begin{tabular}{lccc}
                \hline\hline            
                Star & Mass & Radius \\ 
                 & $(M_\odot)$ & $(R_\odot)$ \\
                \hline
                $\alpha$\,Cen\,A & $\massA \pm \massAerr$$^a$ & $\radiusA \pm \radiusAerrtot$$^b$ \\
                $\alpha$\,Cen\,B & $\massB \pm \massBerr$$^a$ & $\radiusB \pm \radiusBerrtot$$^b$ \\
                $\alpha$\,Cen\,A+B & $\massAB \pm \massABerr$$^a$ & $-$ \\
                Proxima & $\mProxMann \pm \mProxMannerr$$^c$ & $\RProxMann \pm \RProxMannerr$$^c$ \\
        \hline
        \end{tabular}
        \tablefoot{$^a$ \citetads{2016kervella}, $^b$ \citetads{2016bkervella}, $^c$ \citetads{0004-637X-804-1-64}.}
\end{table}

\section{Dynamics of the $\alpha$\,Cen--Proxima system\label{astrometry}}

\subsection{Astrometry, proper motions and parallaxes}

\begin{table*}[ht]
        \caption{Positional data, parallax and radial velocity of $\alpha$\,Centauri AB (barycenter) and Proxima.
        The coordinates are expressed in the ICRS for the \emph{Hipparcos} epoch (1991.25).
        The barycentric radial velocity $v_\mathrm{r,\,abs}$ is corrected for the convective blueshift and gravitational redshift.}
        \label{posveldata}
        \centering
        \renewcommand{\arraystretch}{1.2}
        \begin{tabular}{lcccccc}
                \hline\hline
                Object & $\alpha \pm \sigma_\alpha$ & $\delta \pm \sigma_\delta$ & $\pi$ & $\mu_\alpha$ & $\mu_\delta$ & $v_\mathrm{r,\,abs}$ \\ 
                 & (h:m:s $\pm$ mas) & (d:m:s $\pm$ mas) & (mas) & (mas\,a$^{-1}$) & (mas\,a$^{-1}$) & (km\,s$^{-1}$) \\
                \hline          \noalign{\smallskip}
                $\alpha$\,Cen & 14:39:40.2068 $\pm 25$$^a$ & -60:50:13.673 $\pm 19$$^a$ & $\parallaxAlfCen \pm \parallaxAlfCenerr$$^b$ & $\barypmRA \pm \barypmerr$$^b$ & $\barypmdec \pm \barypmerr$$^b$ & $\RVAlfCen \pm \RVAlfCenerr$$^b$ \\
                Proxima & 14:29:47.7474 $\pm 1.3$$^a$ & -62:40:52.868 $\pm 1.5$$^a$ & $\parallaxProx \pm \parallaxProxerr$$^c$ &$\ProxpmRA \pm \ProxpmRAerr$$^c$ & $\Proxpmdec \pm \Proxpmdecerr$$^c$ & $\RVProx \pm \RVProxerr$$^d$ \\
                \hline
        \end{tabular}
        \tablebib{
        $^a$ \citetads{1997ESASP1200.....E};
        $^b$ \citetads{2016kervella};
        $^c$ \citetads{1999AJ....118.1086B};
        $^d$ Present work.
        }
\end{table*}

We adopt the position of the barycenter of $\alpha$\,Cen determined by \citetads{2016kervella} at the \emph{Hipparcos} epoch (1991.25), and the corresponding position of Proxima also from \emph{Hipparcos}.
The presence of occasional flares in Proxima is not expected to significantly
affect its apparent position \citepads{1998ASPC..154.1212B}.
The parallax of $\alpha$\,Cen is taken from \citetads{2016kervella} and that of Proxima is adopted from \citetads{1999AJ....118.1086B}, whose value is compatible with the measurement by \citetads{2014AJ....148...91L} ($\pi = 768.13 \pm 1.04$\,mas).
An overview of the astrometric parameters is presented in Table~\ref{posveldata}.
Proxima is closer to us than $\alpha$\,Cen by $44.8 \pm 1.5$ light-days.
This implies that Proxima's position on sky in the $\alpha$\,Cen time referential is shifted by $(+0.46\arcsec, -0.09\arcsec)$.
We applied this correction to the apparent position of Proxima but the effect is negligible on the derived orbital parameters.
However, it will be necessary to consider it to interpret the coming \emph{Gaia} \citepads{2016arXiv160904153G} observations of Proxima.
The linear separation between the barycenter of $\alpha$\,Cen\,AB and Proxima is $d_{\alpha-\mathrm{Prox}} = \distABProx \pm \distABProxerr$\,au.
We neglect the change in differential RV between the time of the astrometric measurement by \emph{Hipparcos} and the mean epoch of the HARPS spectra of Proxima ($\mathrm{MJD} \approx 56100$).
This is justified by the fact that no secular acceleration has been detected in Proxima.
We also neglect the transverse Doppler redshift predicted by the special relativity theory (lower than 1\,m\,s$^{-1}$ and identical for $\alpha$\,Cen and Proxima).
From the coordinates, parallax, proper motion and RVs of $\alpha$\,Cen and Proxima, we compute their 3D solar-centric positions $(X,Y,Z)$ and their heliocentric Galactic space velocity vectors $(U,V,W)$ (Table~\ref{spacevel}).

\subsection{Orbital parameters\label{orbit}}

The relative velocity above which Proxima would not be gravitationally bound to $\alpha$\,Cen is
\begin{equation}
v_\mathrm{max} = \sqrt{\frac{2\,G\,m_\mathrm{tot}}{d_{\alpha-\mathrm{Prox}}}} = 545 \pm 11\,\mathrm{m\,s}^{-1},
\end{equation}
where $m_\mathrm{tot} = m_A + m_B + m_\mathrm{Proxima} = 2.165 \pm 0.008\,M_\odot$ (Table~\ref{physicalparameters}).
The difference of the space velocity vectors of $\alpha$\,Cen\,AB and Proxima has a norm of $v_\mathrm{\alpha-Prox} = \velprox\,\mathrm{m\,s}^{-1}$.
The observed velocity is therefore lower than the unbound velocity limit by $-5.4\,\sigma$, corresponding to a theoretical probability of $4 \times 10^{-8}$ that the stars are not gravitationally bound.
This conclusion is robust with respect to the adopted GRS correction; the adoption of the interferometric radius value for Proxima \citepads{2009A&A...505..205D} instead of the predicted value from \citetads{0004-637X-804-1-64} results in a velocity of $v_\mathrm{\alpha-Prox} = 309 \pm 55\,\mathrm{m\,s}^{-1}$ and a $-4.2\,\sigma$ difference with the unbound velocity value.

\begin{table}[]
        \caption{Orbital parameters of Proxima.}
        \label{orbitalparameters}
        \centering
        \renewcommand{\arraystretch}{1.3}
        \begin{tabular}{lll}
                \hline\hline            
                Parameter & Value & Unit\\ 
                \hline          \noalign{\smallskip}
 Semi-major axis $a$ & $\sma$ & kau \\ 
 Excentricity    $e$ & $\ecc$ & \\ 
 Period          $P$ & $\Per$ & ka \\ 
 Inclination     $i$ & $\incl$ & $\deg$ \\ 
 Longitude of asc. node $\Omega$   & $\Omegaval$ & $\deg$ \\ 
 Argument of periastron $\omega$   & $\omegaval$ & $\deg$ \\ 
 Epoch of periastron $T_0$$^a$ & $\Tzero$ & ka \\ 
  \noalign{\smallskip} \hline 
 Periastron radius & $\Periastron$ & kau \\ 
 Apastron radius   & $\Apastron$ & kau \\ 
      \noalign{\smallskip}              \hline
        \end{tabular}
        \tablefoot{$^a$ The epoch of periastron passage $T_0$ is relative to present.}
\end{table}

\begin{figure}
        \centering
        \includegraphics[width=7.5cm]{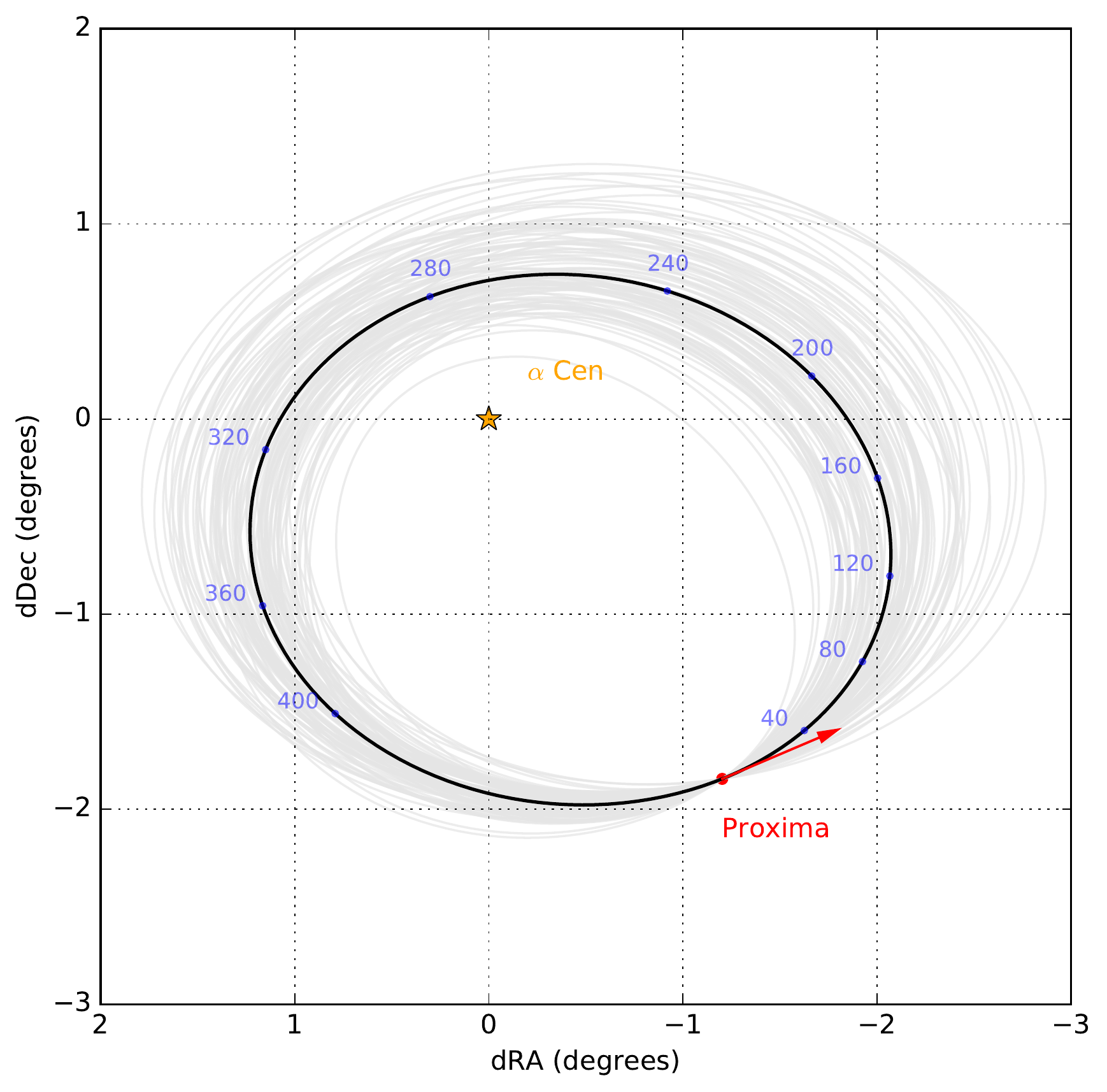}
        \caption{Best-fit orbital trajectory of Proxima around $\alpha$\,Cen projected in the plane of the sky (black curve), with a set of 100 possible orbits computed using a Monte Carlo approach (thin grey curves).
        The blue dots represent the position of Proxima every 40\,000 years (labels in millenia from present) and the direction of the present velocity vector of Proxima is shown as a red arrow.
        \label{sky-plot}}
\end{figure}

\begin{figure}
        \centering
        \includegraphics[width=7.5cm]{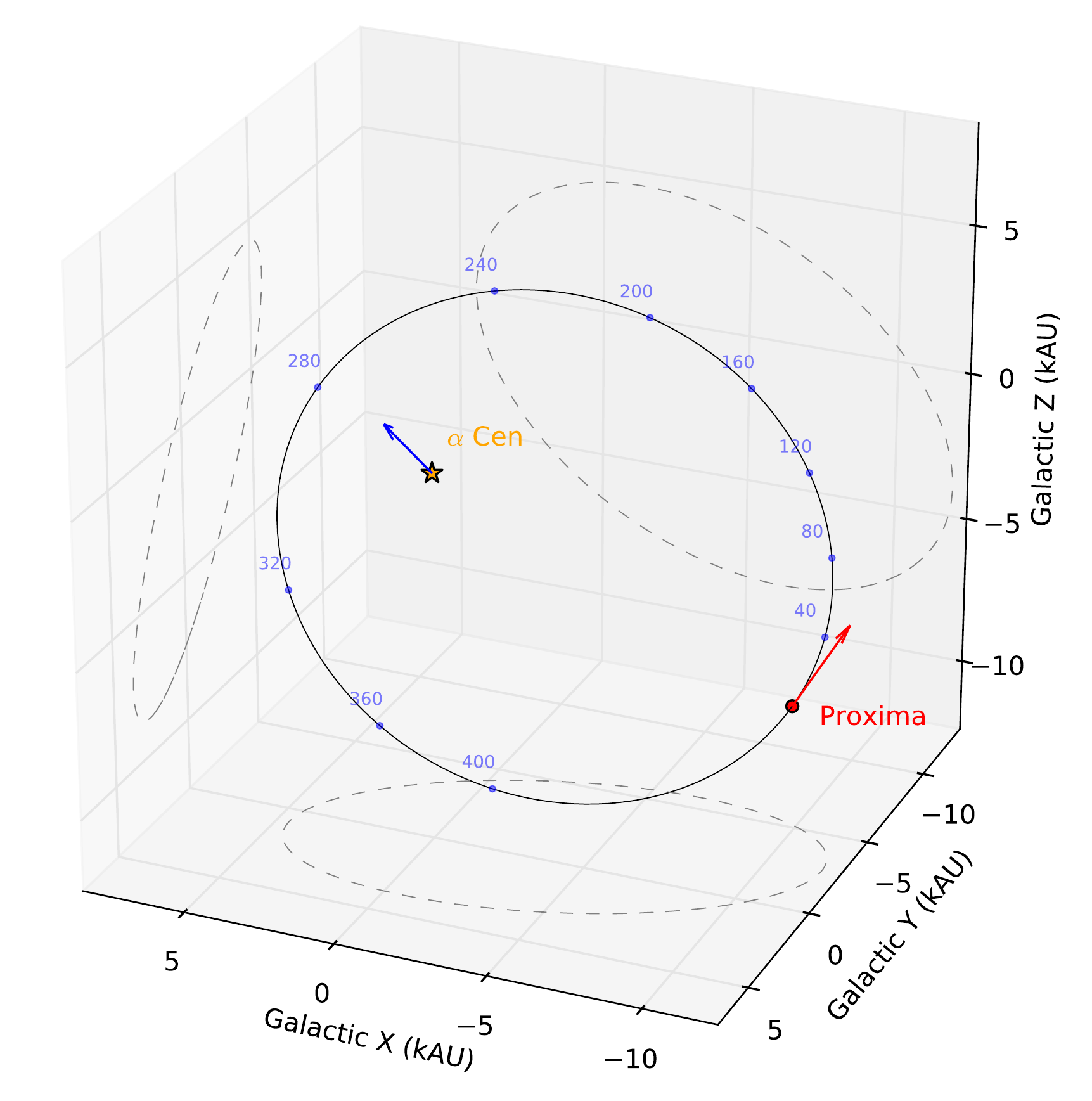}
        \caption{Orbit in 3D cartesian Galactic coordinates, with the Sun-$\alpha$\,Cen direction shown as a blue vector originating from $\alpha$\,Cen.
        The labels are the same as in Fig.~\ref{sky-plot}.
        \label{3D-plot}}
\end{figure}

\begin{figure}[ht]
        \centering
        \includegraphics[width=7.5cm]{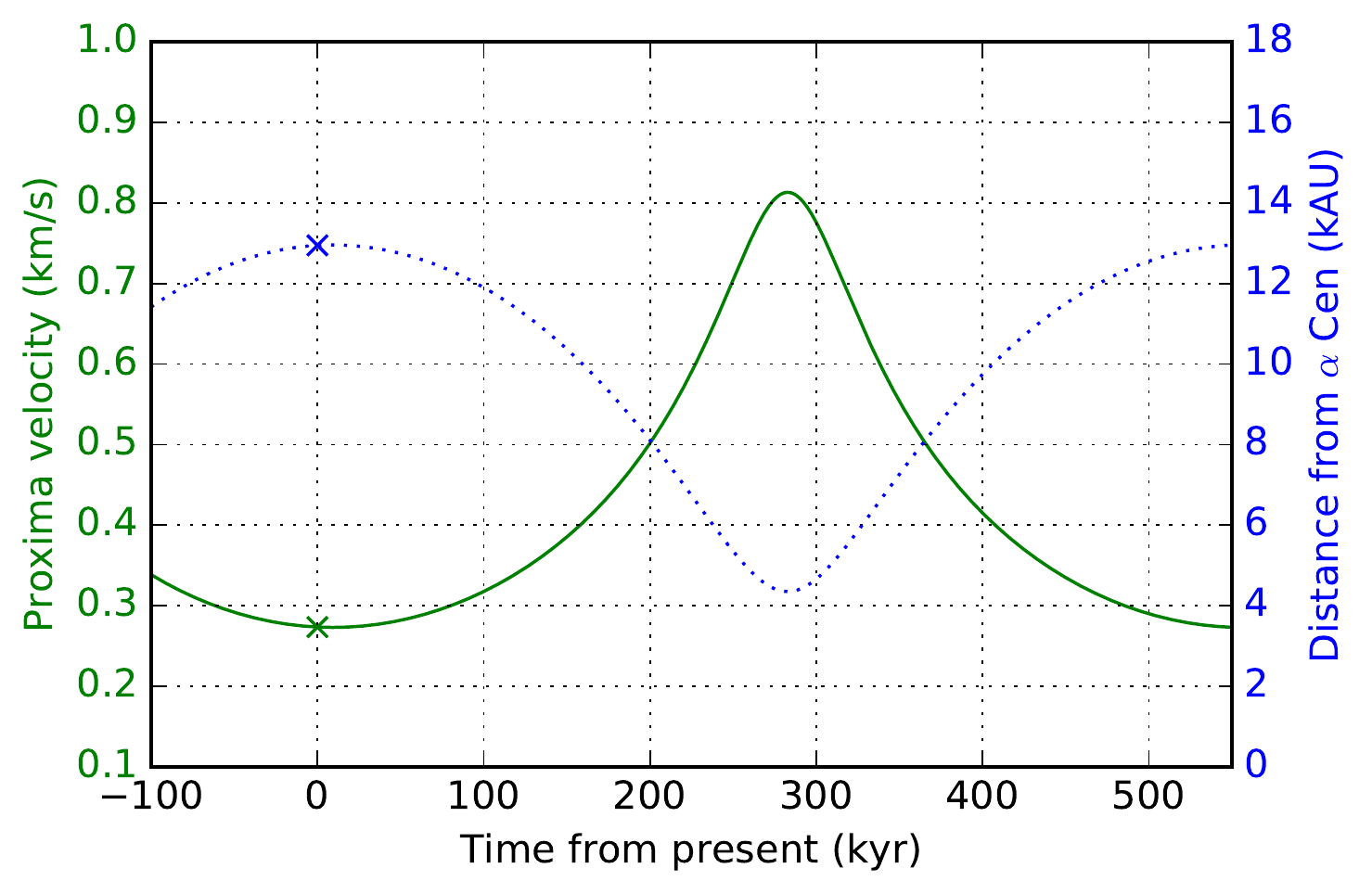}
        \caption{Velocity (solid line) and separation (dotted line) of Proxima relative to $\alpha$\,Cen. The present values are marked with crosses.
        \label{velocityradius}}
\end{figure}

We computed the orbital parameters of Proxima (Table~\ref{orbitalparameters}) using, as inputs, the total mass of the system, the 3D position and the 3D Galactic space velocity of Proxima with respect to the barycenter of $\alpha$\,Cen ($m_\mathrm{tot},X,Y,Z,U,V,W$).
The error bars were derived using a classical Monte Carlo approach.
We drew a large number (100\,000) of sets of input measurements with random fluctuations according to their error bars.
The corresponding sets of orbital parameters were computed and the error bars were obtained from the 16th and 84th percentiles of their histograms (68\% confidence interval; Fig.~\ref{HistoOrbit}).
It is interesting to remark that the derived parameters are qualitatively similar to the range of possible values found by \citetads{1966AJ.....71.1017G}.
The orbit of Proxima is represented in projection on the plane of the sky in Fig.~\ref{sky-plot} and in cartesian Galactic coordinates in Fig.~\ref{3D-plot}.
Figure~\ref{velocityradius} shows the velocity and separation of Proxima with respect to $\alpha$\,Cen over its orbit.
The orbital plane of Proxima is inclined by $\approx 30^\circ$ with respect to that of $\alpha$\,Cen AB ($i = 79^\circ$; \citeads{2016kervella}).

\section{Conclusions}
Using high-accuracy RV measurements and astrometry, we show with a high level of confidence that Proxima is gravitationally bound to $\alpha$\,Cen and orbits the pair on a moderately eccentric, very long-period orbit.
This conclusion is particularly valuable for the modeling of this star as it means that the three stars are coeval and share the same initial metallicity.
Due to the very weak gravitational interaction between Proxima and $\alpha$\,Cen, \citetads{2009MNRAS.399L..21B, 2011Ap&SS.333..419B, beech2015} proposed that this system could be a test case for the modified Newtonian dynamics (MOND) theory \citepads{1983ApJ...270..365M, 2004PhRvD..70h3509B}.
Such a wide multiple system may have formed during the dissolution of their original star cluster \citepads{2010MNRAS.404.1835K}.
In spite of its large semi-major axis, the statistical dissolution time of the Proxima-$\alpha$\,Cen system is expected to be much longer than 10\,Ga \citepads{ 2010MNRAS.401..977J,1985ApJ...290...15B}.
The orbital motion of Proxima could have played a role in the formation and evolution of its planet \citepads{2016Natur.536..437A}.
Conversely, it may also have influenced circumbinary planet formation around $\alpha$\,Cen \citepads{2016arXiv160703090W}.
A speculative scenario is that \object{Proxima b} formed as a distant circumbinary planet of the $\alpha$\,Cen pair, and was subsequently captured by Proxima.
Proxima b could then be an ocean planet resulting from the meltdown of an icy body \citepads{2016arXiv160909757B}.
This would also mean that Proxima b may not have been located in the habitable zone \citepads{2016arXiv160806813R} for as long as the age of the $\alpha$\,Cen system (5 to 7\,Ga; \citeads{2005A&A...441..615M}; \citeads{2004A&A...417..235E}; \citeads{2003A&A...404.1087K}; \citeads{thevenin02}).

\begin{acknowledgements}
We are grateful to Dr James Jenkins for discussions that led to important improvements of this Letter.
CL acknowledges the financial support of the Swiss National Science Foundation (SNSF).
This research made use of Astropy\footnote{\url{http://www.astropy.org/}} \citepads{2013A&A...558A..33A}, and of the SIMBAD \citepads{2000A&AS..143....9W}, VIZIER (CDS, Strasbourg, France) and NASA's Astrophysics Data System databases.
\end{acknowledgements}

%
\bibliographystyle{aa} 
\bibliography{BiblioAlfCen.bib} 
%



%

\begin{appendix}

\section{Radial velocity of Proxima \label{radvelProxima}}

\begin{figure*}[ht]
        \centering
        \includegraphics[width=\hsize]{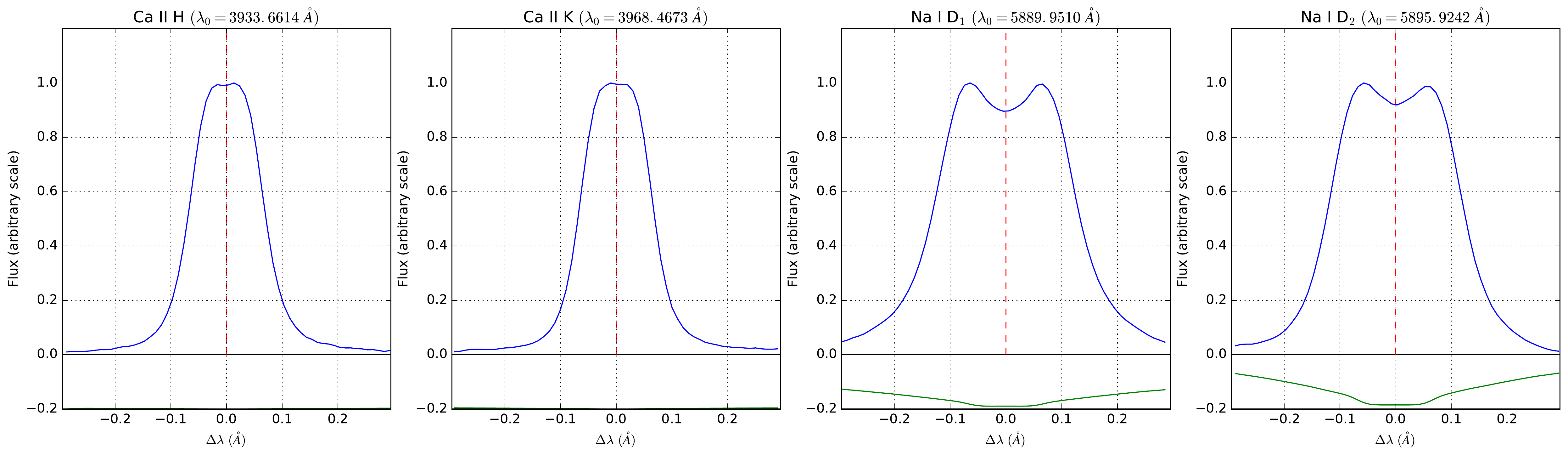}
        \caption{Average profiles of the selected emission lines of Proxima (blue curves), with the subtracted background level (green curves).
        The dashed red line marks the wavelength of the lines at a velocity of  $v = \RVProxObs\,\mathrm{km\,s}^{-1}$.
        \label{velocity-lines}}
\end{figure*}

The main difficulties to determine the absolute RV of red dwarfs from cross correlation with model templates is the extremely strong line blending, and our incomplete knowledge of molecular line transitions.
The high resolution spectra produced by recent atmosphere models (e.g., PHOENIX, \citeads{2013A&A...553A...6H}) provide a sufficient accuracy to determine the basic parameters of the stars, but they fail to reproduce many of the observed lines and the details of the line profiles.
In addition, the wavelengths of the molecular lines are in general not known with a sufficient accuracy to reach the $\mathrm{m\,s}^{-1}$ level.
To overcome these limitations, we selected four strong very high signal-to-noise emission lines of Ca II and Na I, whose wavelengths are very accurately known \citepads{2008PhRvA..78c2511W, 2003ApJS..149..205M,2004ApJS..151..403M}.
We subtracted the spectral background at the position of these emission lines from a PHOENIX\footnote{\url{http://phoenix.astro.physik.uni-goettingen.de}} model spectrum \citepads{2013A&A...553A...6H} that was scaled to the flux of the HARPS spectrum over a neighboring emission-free region.
We checked that alternate methods to estimate the background level (linear, constant) do not affect the derived RV at a $\pm 20$\,m\,s$^{-1}$ level.
Removing one of the lines from the sample also does not lead to biases beyond this level.
We obtained the Doppler shift of each emission line separately from the measurement of the barycenter of their emission over a $\Delta \lambda = \pm 0.2\,\AA$ region (Fig.~\ref{velocity-lines}).

\begin{figure}[ht]
        \centering
        \includegraphics[width=8cm]{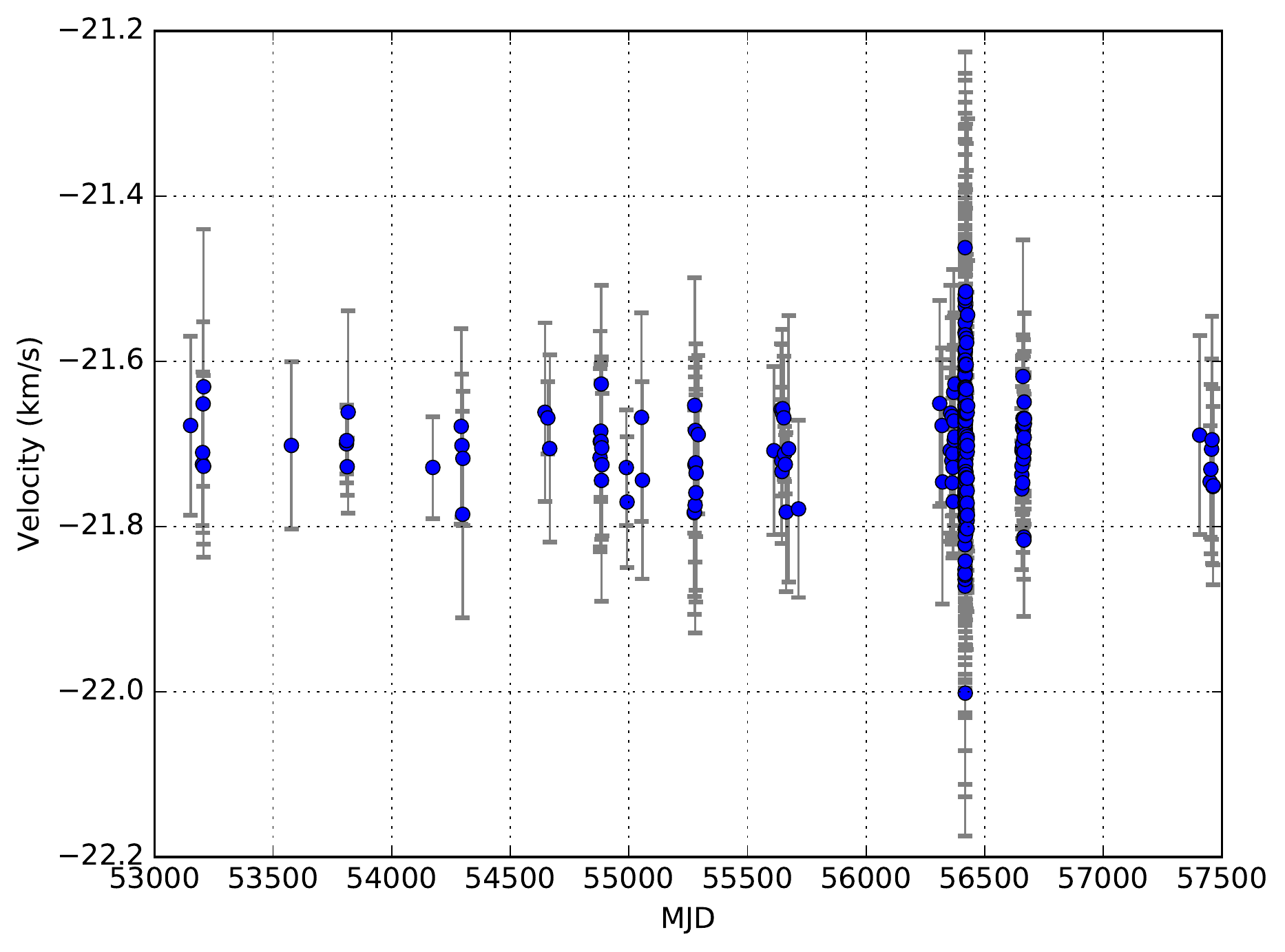}
        \caption{Times series of the measured absolute radial velocities of Proxima over a period of 12 years.
        \label{RVtime}}
\end{figure}

\begin{figure}[ht]
        \centering
        \includegraphics[width=8cm]{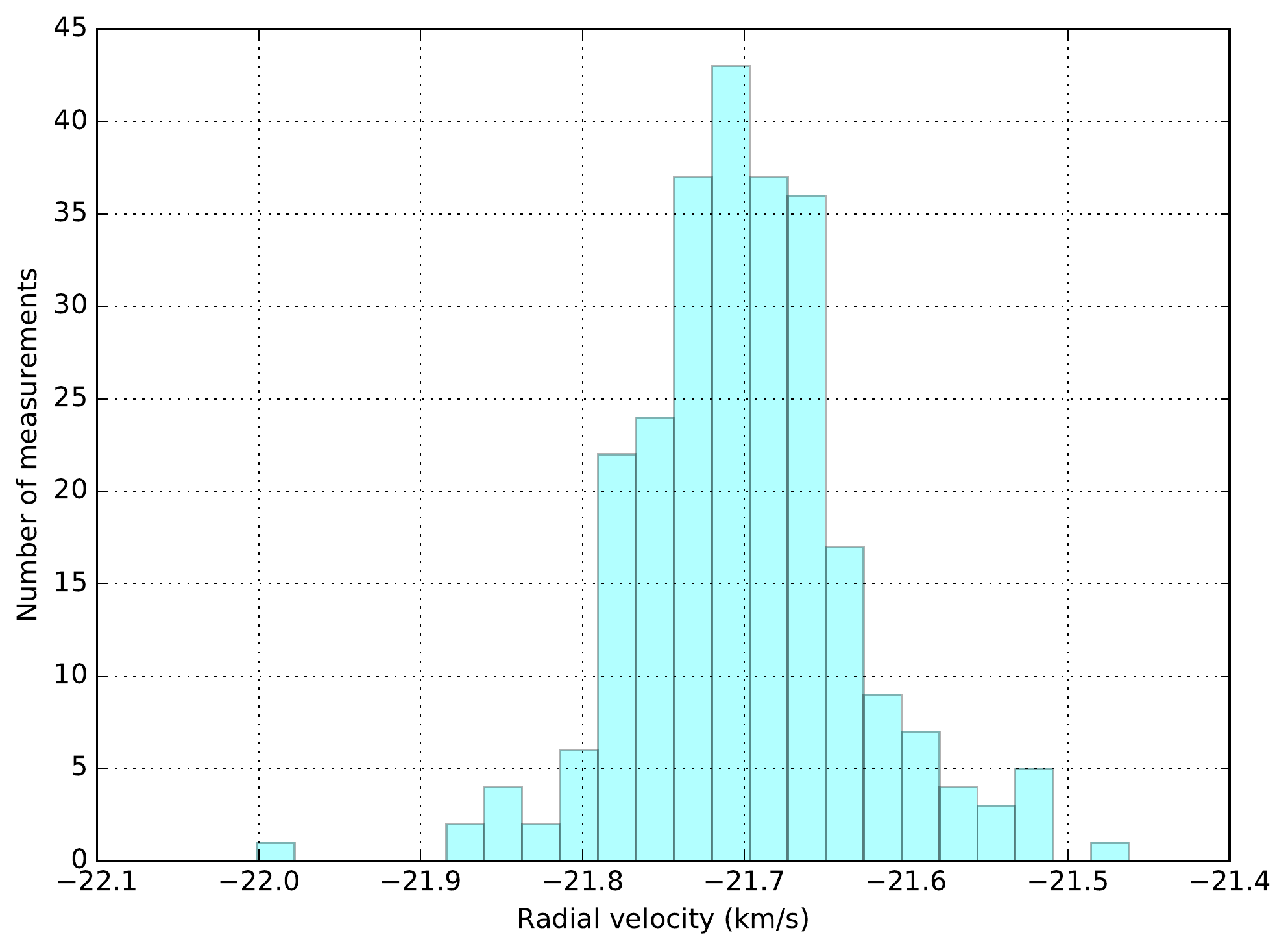}
        \caption{Histogram of the measured absolute radial velocities of Proxima.
        \label{RVhisto}}
\end{figure}

Over the 271 HARPS spectra of Proxima present in the ESO Phase 3 archive, we kept 260 that provide a good consistency between the velocities estimated using the four emission lines (within 300\,m\,s$^{-1}$).
The velocities of the four lines were averaged to obtain one measurement per HARPS epoch, whose time sequence is represented in Fig.~\ref{RVtime}.
The standard deviation of all epoch measurements is $\sigma = 0.068$\,km\,s$^{-1}$, and the histogram of the measurements is shown in Fig.~\ref{RVhisto}.
We used a bootstrapping approach to estimate the statistical uncertainty of the resulting velocity ($\pm \RVProxObserrstat$\,km\,s$^{-1}$).
We add quadratically a $\pm 0.025$\,km\,s$^{-1}$ systematic uncertainty to account for the background and line selection dispersion.
We thus obtain the barycentric velocity measure \citepads{2003A&A...401.1185L} of Proxima $v_\mathrm{r,\,obs}[\mathrm{Proxima}] = \RVProxObs \pm \RVProxObserrstat \pm 0.025\,\mathrm{km\,s}^{-1}$

\section{3D positions and velocities}

The heliocentric Galactic coordinates of the barycenter of $\alpha$\,Cen and of Proxima are presented in Table~\ref{spacevel}, together with their heliocentric space velocity.
The differential position and velocity vectors between $\alpha$\,Cen and Proxima are also listed.
We followed the classical convention of
$X$ and $U$ increasing toward the Galactic center,
$Y$ and $V$ positive toward the Galactic direction of rotation, and
$Z$ and $W$ positive toward the North Galactic pole.

\begin{table*}
        \caption{Heliocentric coordinates $(X,Y,Z)$ and space velocity vectors $(U, V, W)$ of $\alpha$\,Cen and Proxima in the Galactic frame. 
        }
        \label{spacevel}
        \centering
        \renewcommand{\arraystretch}{1.2}
        \begin{tabular}{lccc}
                \hline\hline            \noalign{\smallskip}
                Parameter & $\alpha$\,Cen & Proxima &$\mathrm{Proxima} - \alpha$\,Cen\\ 
                \hline          \noalign{\smallskip}
                $X$ (pc) & $+0.95845 \pm 0.00078$ & $+0.90223 \pm 0.00043$ &  $-0.05622 \pm 0.00089$ \\
                $Y$ (pc) & $-0.93402 \pm 0.00076$ & $-0.93599 \pm 0.00045$ & $-0.00198 \pm 0.00089$  \\
                $Z$ (pc) & $-0.01601 \pm 0.00001$ & $-0.04386 \pm 0.00002$ &  $-0.02785 \pm 0.00002$ \\
                \hline          \noalign{\smallskip}
                $U$ (km\,s$^{-1}$) & $-29.291 \pm 0.026$ & $-29.390 \pm 0.027$ &  $-0.099 \pm 0.038$ \\
                $V$ (km\,s$^{-1}$) & $+1.710  \pm 0.020$ & $+1.883 \pm 0.018$ & $+0.173 \pm 0.027$ \\
                $W$ (km\,s$^{-1}$) & $+13.589 \pm 0.013$ & $+13.777 \pm 0.009$ &  $+0.187 \pm 0.016$ \\
                \hline
        \end{tabular}
\end{table*}

\section{Orbital parameters statistics}

\begin{figure*}[ht]
        \centering
        \includegraphics[width=\hsize]{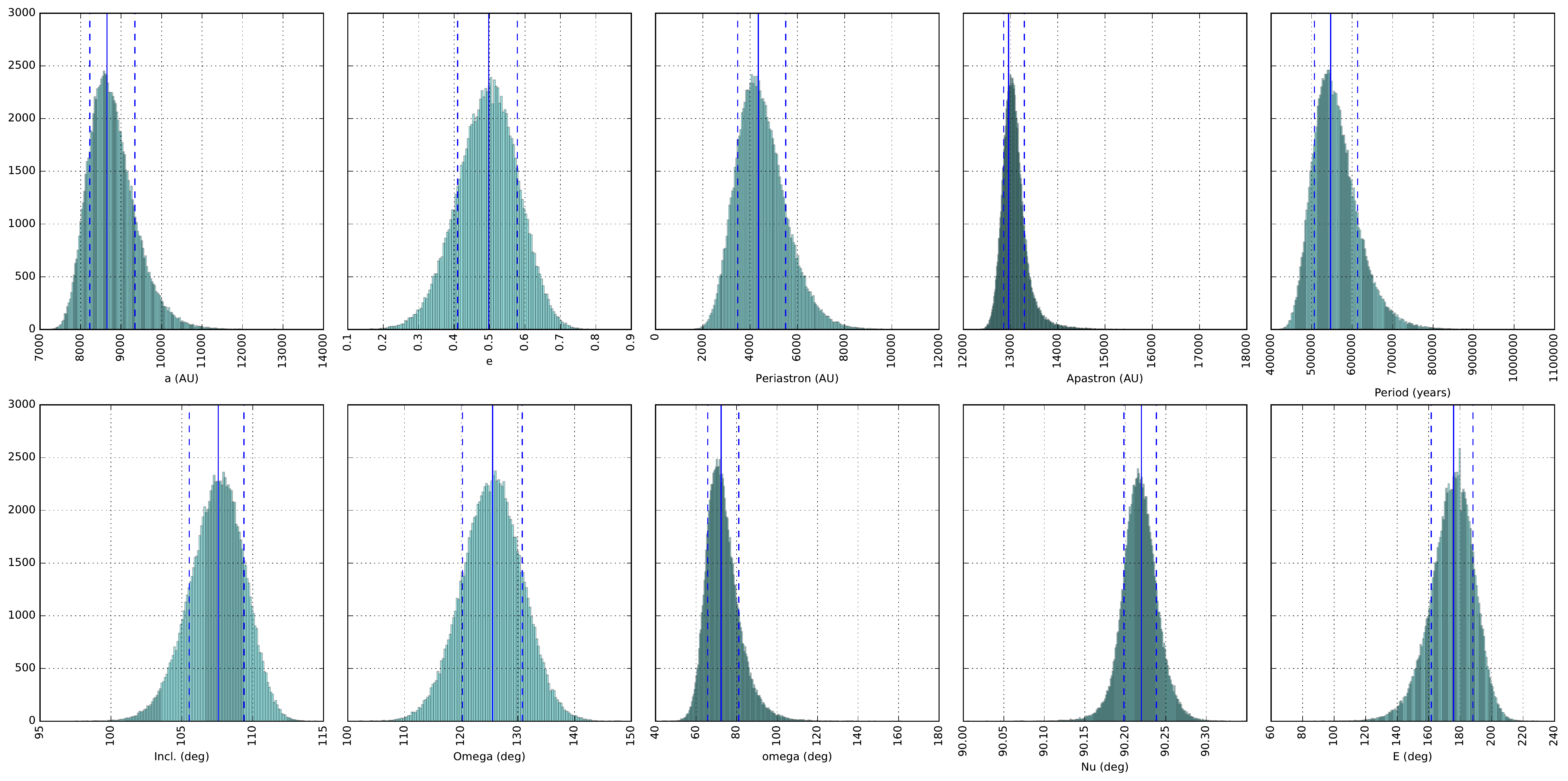}
        \caption{Histograms of the Monte Carlo simulations of the orbital parameters of Proxima. The solid blue line is the best fit parameter, and the dashed lines delimit the 68\% confidence interval ($1\sigma$).
        \label{HistoOrbit}}
\end{figure*}

The histograms of the values of the orbital parameters of Proxima from our Monte Carlo simulations is presented in Fig.~\ref{HistoOrbit}.

\end{appendix}

\end{document}